\documentclass{jpconf}

\usepackage{amsmath}
\usepackage{amssymb}
\usepackage{mathtools}
\usepackage{tikz}
\usepackage{graphicx}
\usepackage{caption}
\usepackage{hyperref}
\usepackage{subcaption}
\usepackage[utf8]{inputenc}
\usepackage{xcolor}
\usepackage[toc]{appendix}

\begin{document}

\title{Thermodynamic Stability of Schwarzschild-de Sitter Black holes with R\'enyi entropy}

\author{Thanawat Anusonthi$^{1}$, Pitayuth Wongjun$^{1}$ and Ratchapat Nakarachinda$^{1,2}$}

\address{$^1$ The Institute for Fundamental Study (IF), Naresuan University, 99 Moo 9, Tah Poe, Mueang Phitsanulok, Phitsanulok, 65000, Thailand} 

\address{$^2$ Department of Mathematics and Computer Science, Faculty of Science, Chulalongkorn University, 254 Phaya Thai Rd., Pathum Wan, Bangkok, 10330, Thailand}

\ead{anusonpun@gmail.com}

\begin{abstract}
Even though, classically, a black holes is gravitational object, it can be treated as a thermodynamic object when quantum effect is taken into account. It is found that by using Gibbs-Boltzmann entropy, thermodynamic system associated Schwarzschild-de Sitter black hole is unstable while it is stable under description of  R\'enyi entropy. According to R\'enyi entropy,  the thermodynamic phase space needed to be extended. Specifically, the non-extensive parameter must be treated as a thermodynamic variable. In this work, we investigate the thermodynamic stability of Schwarzschild-de Sitter black hole with R\'enyi entropy by treating non-extensive parameter as chemical potential. We found that it is possible to obtain the stability of the black hole under the process with fixing pressure, temperature and number of particles. We also found that there exist the phase transitions from the hot gas to the stable black hole in such the process. Therefore, if such a black hole is possibly observed, the non-extensive effect of R\'enyi statistics may provide a physical insight beyond the standard approach to black hole thermodynamics.
\end{abstract}

\section{Introduction \& Motivation}
\ \ \ After Hawking temperature and Bekenstein-Hawking entropy were explored \cite{Re1,Re2}, a black hole can be treated as a thermodynamic object. Thus, the stability of a black hole in thermodynamic aspect is worthwhile to investigate. By considering heat capacity and Gibbs free energy, one found that the Schwarzschild black hole is thermodynamically unstable \cite{Re5}. The Hawking temperature can be obtained by computing the spectra of emitted particles and comparing them with one from black-body radiation. This procedure is based on Gibbs-Boltzmann statistics. Since the entropy of the black hole is proportional to area of the horizon, such entropy is nonextensive. Then, Gibbs-Boltzmann statistics in which the entropy is extensive quantity may not be used to investigate the thermodynamics of the black hole. Therefore, it might be appropriate to study the thermodynamic stability of a black hole using another type of entropy. One of the generalized entropies widely used is Tsallis entropy \cite{Re6,Re7}. It is worthwide to reinterpret the horizon area as Tsallis entropy. 
Since thermodynamics is usually studied in the context of thermodynamic equilibrium, such the corresponding  entropy must satisfy the 0th law of thermodynamics. However, Tsallis entropy does not satisfy the 0th law. One of the possible treatments for such issue is using logarithmic map of Tsallis entropy. This kind of mapped entropy is known as ``R\'enyi entropy". Therefore, we would investigate the black hole's thermodynamics by using R\'enyi entropy \cite{Re7}. \\ 
\par In order to investigate the proper thermodynamics of black holes, one needs a suitable 1st law of black holes' thermodynamics. This law can be obtained from considering of Smarr's formula via Euler's theorem of a homogeneous function. To obtain the Smarr's formula and the 1st law, we have to treat the non-extensive parameter for R\'enyi entropy as a thermodynamic variable. This method is referred to ``phase-space extension" \cite{Re9,Re10,Re11}. \\ 
\par According to observations, the Universe is expanding with acceleration \cite{Re15,Re16}. One of the possible ways to describe this phenomenon is introducing the cosmological constant into Einstein's field equation. A static \& spherically symmetric black hole solution of this equation is known as ``Schwarzschild-Tangherlini black hole". The Schwarzschild-Tangerlini black hole has two branches depending on a sign of the cosmological constant. For the positive (negative) sign of the cosmological constant, the black hole is called ``Schwarzschild-de Sitter (anti de Sitter) black hole". It has been found that 
the Schwarzschild-de Sitter black hole is thermodynamically unstable by using Gibbs-Boltzmann statistics, and it is possible to obtain a thermodynamically stable black hole by using R\'enyi statistics \cite{Re5,Re18}. However, such investigations adopts the Smarr's formula and the 1st law. Then, it is not consistent with geometric relation of black holes. As we mentioned before, to obtain suitable Smarr's formula via R\'enyi entropy, one needs to promote the nonextensive parameter to a thermodynamic variable. By interpreting the nonextensive parameter as chemical potential, it is possible to consider the thermodynamic process with fixing the number of particles. In this work, we will show that the Schwarzschild-de Sitter black hole can be stabilized according to R\'enyi entropy. Moreover, the evolution of the black hole will be analyzed via thermodynamic stability. \\ 

In this work, we use Planck's unit system and organize as follows. In Section \ref{bhwre}, the formulation of R\'enyi and Tsallis entropy of a black hole's entropy is introduced. Determining Smarr's formula for Schwarzschild-de Sitter black hole and the representation of the black hole's mass in thermodynamics is done in Section \ref{thmsch}. The stability analyses and calculations for the black hole are shown in Section \ref{cal}. In section \ref{sum}, we summarize and discuss the results.

\section{Black holes with R\'enyi entropy} \label{bhwre}
\ \ \  By comparing black body's radiation based on Gibbs-Boltzmann statistics, Hawking obtained the temperature of a black hole called Hawking temperature \cite{Re1}. Thus, the corresponding entropy which known as the Bekenstein-Hawking (BH) entropy \cite{Re2} can be expressed as
\begin{align}
S_{BH} = \frac{A}{4} ,
\end{align}
where $A$ is the surface area at the black hole's event horizon. Obviously, the Bekenstein-Hawking entropy is non-additive and corresponds to non-extensive entropy. Meanwhile, the Gibbs-Boltzmann entropy is additive and extensive, it should be more appropriate if the Bekenstein-Hawking entropy is represented by other non-extensive entropy. One can interpret the black hole's entropy as Tsallis entropy which is 1-parameter generalized entropy \cite{Re5,Re11} as follows
\begin{equation}
S_{BH} = S_{GB} \quad \rightarrow \quad S_{BH} = S_T. \nonumber
\end{equation}

Based on thermodynamic equilibrium, the entropy of the system must be additive entropy in order to be consistent with the 0th law of thermodynamics \cite{Re13}. In fact, the additive entropy can be obtained by taking the logarithmic map of Tsallis entropy \cite{Re7}. The resulting entropy known as R\'enyi entropy \cite{Re14} is expressed by
\begin{equation}
S_R = \frac{1}{\lambda} \ln{(1+\lambda S_{T})} = \frac{1}{\lambda} \ln{\left( \sum_i p_i^{1-\lambda} \right)},
\end{equation}
where $\lambda$ is ``non-extensive parameter" viable in the range $(-\infty,1)$. In the limit $\lambda \rightarrow 0$, the R\'enyi entropy recovers Gibbs-Boltzmann entropy,
\begin{align} \label{limsr}
\lim_{\lambda \rightarrow 0} S_R = S_{GB}.
\end{align} 
Then, the entropy of a black hole can be written in terms of R\'enyi entropy as
\begin{equation} \label{sbh}
S_{bh} = \frac{1}{\lambda} \ln{\left( 1+\lambda S_{BH} \right)}.
\end{equation}
The subscript $bh$ stands for ``black hole" rather than ``Bekenstein-Hawking".

\section{Thermodynamic system of Schwarzschild-de Sitter black hole} \label{thmsch}
\ \ \ The Schwarzschild-de Sitter black hole solution can be written as 
\begin{align} \label{metric}
ds^2 = -f(r)dt^2 + f(r)dr^2 + r^2d\Omega^2,
\end{align}
with
\begin{align} \label{fr}
f(r) = 1- \frac{2m}{r} - \frac{\Lambda}{3}r^2 \quad \text{where} \quad \Lambda > 0.
\end{align}
The parameter $m$ is the ADM mass \cite{Re15} and $\Lambda$ is the cosmological constant. At the event horizon, $f(r_H)=0$, the mass of the black hole can be written in terms of the black hole's entropy and cosmological constant as  
\begin{align} \label{mass}
m = \frac{\sqrt{S_{BH}}}{6 \pi^{3/2}} \left( 3\pi-\Lambda S_{BH} \right) = \frac{1}{6 \pi^{3/2} \lambda} \sqrt{ \frac{e^{\lambda S_{bh}} -1}{\lambda} } \left( 3\pi \lambda + \Lambda - e^{\lambda S_{bh}}\Lambda \right),
\end{align}
we have used $S_{BH}= \frac{e^{\lambda S_{bh}}-1}{\lambda}$. The Smarr's formula can be obtained by using property of a homogeneous function with Euler's theorem \cite{Re11}. The Euler's theorem states that if $f(x_1,x_2,...,x_n)$ is a homogeneous function of degree $k$, the function $f(x_1,x_2,...,x_n)$ will satisfy following relations;
\begin{align}
f(\alpha x_1,\alpha x_2,...,\alpha x_n) = \alpha^{k} f(x_1,x_2,...,x_n),  \\
k f(x_1,...,x_n) = \sum_{i=1}^n x_i \frac{\partial f(x_1,...,x_n)}{\partial x_i} , \label{homo}
\end{align}
where $\alpha$ is a real constant. Treating the mass function in Eq.~\eqref{mass} as the homogeneous function of $S_{bh},\Lambda^{-1}$ and $\lambda^{-1}$, $m=m(S_{bh}, \Lambda^{-1},\lambda^{-1})$, we obtain
\begin{align}
m \left(\alpha S_{bh},\alpha \Lambda^{-1},\alpha \lambda^{-1} \right) = \frac{\alpha^{1/2}}{6 \pi^{3/2} \lambda} \sqrt{ \frac{e^{\lambda S_{bh}} -1}{\lambda} } \left( 3\pi \lambda + \Lambda - e^{\lambda S_{bh}}\Lambda \right). 
\end{align}
Thus, the mass function is the homogeneous function of degree $1/2$. By applying the relation in Eq.~\eqref{homo}, one would get 
\begin{align} \label{smarr}
\frac{1}{2} m &= \left( \frac{\partial m}{\partial S_{bh}} \right) S_{bh} + \left( \frac{\partial m}{\partial \Lambda^{-1}} \right) \Lambda^{-1} + \left( \frac{\partial m}{\partial \lambda^{-1}} \right) \lambda^{-1},  \nonumber \\
& = \left( \frac{\partial m}{\partial S_{bh}} \right) S_{bh} - \left( \frac{\partial m}{\partial \Lambda} \right) \Lambda - \left( \frac{\partial m}{\partial \lambda} \right) \lambda.
\end{align}
According to Ref.~\cite{Re11}, the conjugate variable of the black hole's entropy is ``R\'enyi temperature", $T_R \equiv \frac{\partial m}{\partial S_{bh}}$. From Eq.~\eqref{smarr}, one can see that  the cosmological constant and the non-extensive parameter play the role of thermodynamic variables. 

\ \ \ For the cosmological constant, its conjugate variable can be expressed as follows
\begin{align} \label{conjL}
\frac{\partial m}{\partial \Lambda} = \frac{1}{6 \pi^{3/2} \lambda} \sqrt{ \frac{e^{\lambda S_{bh}} -1}{\lambda} } \left( 1 - e^{\lambda S_{bh}} \right).
\end{align}
By substituting $\frac{e^{\lambda S_{bh}}-1}{\lambda}=\pi r^2$ in Eq.~\eqref{conjL}, one obtains $\frac{\partial m}{\partial \Lambda}=-\frac{1}{6}r^3$, which is proportional to 3-dimensional volume. Therefore, the conjugate variable of $\Lambda$ can be interpreted as thermodynamic volume by following expression,
\begin{align} \label{volume}
V_\Lambda=\frac{4}{3}\pi r_H^3=-8\pi\left(\frac{\partial m}{\partial \Lambda}\right).
\end{align}
Thus, the cosmological constant is assigned as thermodynamic pressure: $P_\Lambda\equiv-\frac{\Lambda}{8\pi}$. In order to investigate the physical interpretation of the non-extensive parameter and its conjugate variable, $\lambda$ and $\Psi_\lambda \equiv \frac{\partial m}{\partial \lambda}$, we do the Legendre transformation of the mass function as follows
\begin{align}
E & = m - P_\Lambda V_\Lambda - \Psi_\lambda \lambda,
\end{align}
where $E$ is a thermodynamic potential. The derivative form of $E$ is written by
\begin{align}
dE = T_R dS_{bh} - P_\Lambda dV_\Lambda - \lambda d \Psi_\lambda . \label{dE}
\end{align}
Since the energy E is a function of $S_{bh},V_\Lambda,\Psi_\lambda$, $E=E(S,V,\Psi_\lambda)$. It is reasonable to interpret $E$ as internal energy. Therefore, one can interpret $\Psi_\lambda$ as a number of particles and $\lambda$ will play the role of chemical potential. In fact, by using Taylor's series expansion for small $\lambda$, $\Psi_\lambda$ can be approximated as
\begin{align}
\Psi_\lambda \approx \frac{1}{8}\pi r_H^3 \left(1+8\pi r_H^2P_\Lambda\right)
-\frac{1}{24} \pi^2 r_H^5\left(1+8\pi r_H^2P_\Lambda\right)\lambda
+O\left(\lambda ^2\right).
\end{align}
It is seen that the leading order is proportional to $r_H^3$. Since the leading order of $\Psi_\lambda$ is scaled by size of the system, $\Psi_\lambda$ is an extensive variable. This is reasonable to interpret $\Psi_\lambda$ as the number of particles, $\Psi_\lambda = N_\lambda$. In this sense, the non-extensive parameter, $\lambda$, measures how the internal energy changes while there is a transfer of particles in the system. According to Eq.~\eqref{dE}, the chemical potential can be defined as 
\begin{align}
\mu_\lambda \equiv \left( -\frac{\partial E}{\partial N_\lambda} \right)_{S_{bh},V_\Lambda} = \lambda.
\end{align}
As a result, the 1st law of thermodynamics can be written by
\begin{align} \label{dm}
dm = T_R dS_{bh} + V_\Lambda dP_\Lambda + N_\lambda d\mu_\lambda.
\end{align}

\section{Thermodynamic stability} \label{cal}
\ \ \ In equilibrium thermodynamics, the stability analysis can be divided into two branches, namely local and global stability \cite{Re3,Re4}.
\begin{itemize}
\item Local stability

Local stability is related to the \textit{responsiveness} of a system when we perturb it. The local stability condition is characterized by positive heat capacity,
\begin{align}
C_X = \left( \frac{\delta Q}{\delta T} \right)_X > 0,
\end{align}
where $Q$ stands for heat and the subscript ``$X$" refers to variables that are fixed under the thermodynamic process.  If the system has a negative heat capacity, the temperature of the system will be decreased when the heat transfers into the system by a higher temperature reservoir. Then the more difference in the temperature between the reservoir and the system, the more heat transfer to the system. This leads to the system undergoing far from the equilibrium, opposite to one for positive heat capacity.

\item Global stability

In nature, any system prefers the lowest energy state. Global stability provides us with which state of the thermodynamic system is preferred. In fact, from the relation of entropy and heat in irreversible process $\delta Q \leq T dS$, one can find the change of the Gibbs free energy as $d G \leq -SdT+ V dP -\mu dN$. Considering the thermodynamic process with fixing pressure and number of particles, one obtains $dG \leq -SdT$. As a result, if there are two states with the same temperature, the lower Gibbs free energy state will be preferred to exist in nature. For the hot gas state, there are no horizons or there are no black holes. In fact, all black hole thermodynamic variables are assumed to vanish at this state. As a result, the Gibbs free energy of the hot gas state can be conveniently set to zero. Therefore, by comparing with the hot gas state, the global stability condition on the black hole Gibbs free energy is given by
\begin{align}
\Delta G = G_{\text{black hole}} - G_{\text{hot gas}} = G_{\text{black hole}} < 0.
\end{align}
\end{itemize}

\begin{figure}[h]
\centering
\includegraphics[width=0.5\textwidth]{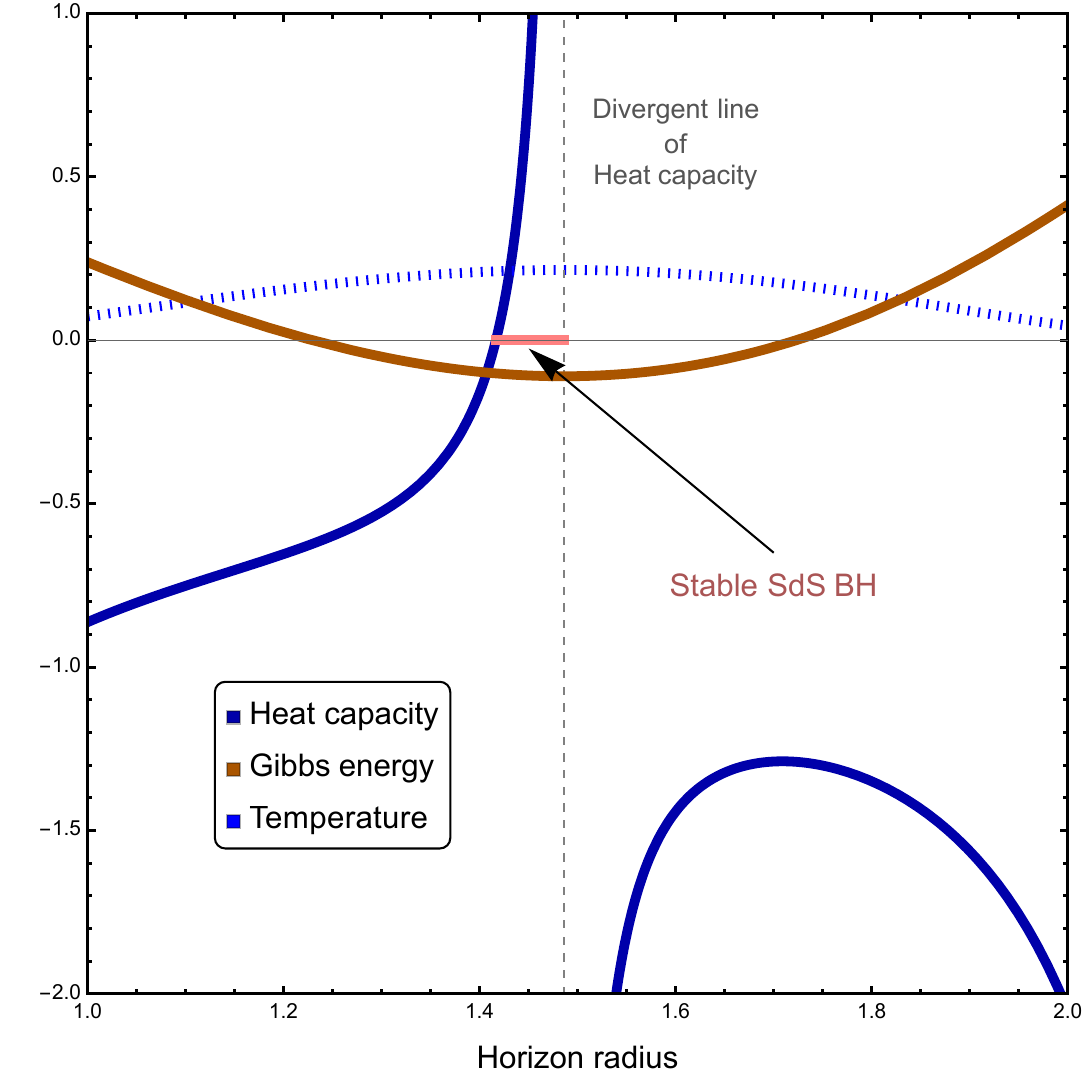}
\caption{Gibbs free energy and heat capacity}
\label{fig:stable}
\end{figure}

In this work, we investigate the thermodynamic stability of the Schwarzschild-de Sitter black hole under the isobaric process of a closed system. In other words, we keep pressure, $\Lambda$, and the conjugate variable of non-extensive parameter, $N_\lambda$, constant. To compute the heat capacity and the Gibbs free energy, we express every quantity as a function of horizon radius, cosmological constant, and non-extensive parameter, $f = f(r_H,\Lambda,\lambda)$. The derivative form of such a function is written by
\begin{align}
\delta f(r_H,\Lambda,\lambda) = \left( \frac{\partial f}{\partial r_H} \right) \delta r_H + \left( \frac{\partial f}{\partial \Lambda}  \right)  \delta \Lambda+ \left( \frac{\partial f}{\partial \lambda} \right) \delta \lambda= \left( \frac{\partial f}{\partial r_H} \frac{\delta r_H}{\delta \lambda}+\frac{\partial f}{\partial \Lambda} \frac{\delta \Lambda}{\delta \lambda} +  \frac{\partial f}{\partial \lambda} \right) \delta \lambda. \label{df}
\end{align}
The factor $\frac{\delta r_H}{\delta \lambda}$ and $\frac{\delta \Lambda}{\delta \lambda}$ are obtained by conditions of considered process. In this paper, we consider the process of fixing pressure and the number of particles. For fixing pressure, since the thermodynamic quantities can be expressed in terms of pressure directly via $ P_\Lambda = -\Lambda/(8\pi)$, it is easy to consider such quantities under the process by fixing $\Lambda$.
For fixing the number of particles, one can find such a condition as follows
\begin{align}
\delta N_\lambda = \left( \frac{\partial N_\lambda}{\partial r_H} \frac{\delta r_H}{\delta \lambda} + \frac{\partial N_\lambda}{\partial \lambda} \right) \delta \lambda = 0 \quad \Rightarrow \quad \frac{\delta r_H}{\delta \lambda} = -  \left. \left( \frac{\partial N_\lambda}{\partial \lambda} \right) \middle/ \left( \frac{\partial N_\lambda}{\partial r_H} \right) \right. . \label{dN}
\end{align}
By considering the Schwarzschild-de Sitter black hole, the heat capacity and the Gibbs free energy can be written as
\begin{align}
C_{P_\Lambda,N_\lambda} = \left. T_R \left( \frac{\delta S_{bh}}{\delta T_R} \right) \right|_{P_\Lambda,N_\lambda} \quad \text{and} \quad G_{bh} = m - T_R S_{bh} - \mu_\lambda N_\lambda. \label{Cpn} 
\end{align}
Substituting the expression for the black hole thermodynamics variables into the above equations and using condition \eqref{dN}, the heat capacity and the Gibbs free energy can be computed explicitly in terms of $r_H, \Lambda, \lambda$. Note that from condition \eqref{dN}, it may not be possible to express $\lambda$ in terms of $N_\lambda$, since the expression contains the complicated term of logarithmic function. In order to obtain the thermodynamic quantities with fixing $N_\lambda$, we use numerical methods. By choosing $\Lambda =0.2$ and $N_\lambda=0.3$, we found that there exists a range of black hole's horizon in which the system is locally and globally stable as illustrated in Figure~\ref{fig:stable}.

From this figure, it shows that the heat capacity diverges at $r=1.486$ which corresponds to the dashed vertical line. The temperature is plotted via the dotted line.  
A small range of horizon radius in which the black hole is thermodynamically stable is represented by the pink highlight. By comparing with the horizon radius, we can track back to the mass of the black hole. For example, a given cosmological constant is $\Lambda=0.2$. The Schwarzschild-de Sitter black hole is thermodynamically stable with a mass around 140 times the mass of the earth.

\begin{figure}[h]
\centering
\includegraphics[width=0.5\textwidth]{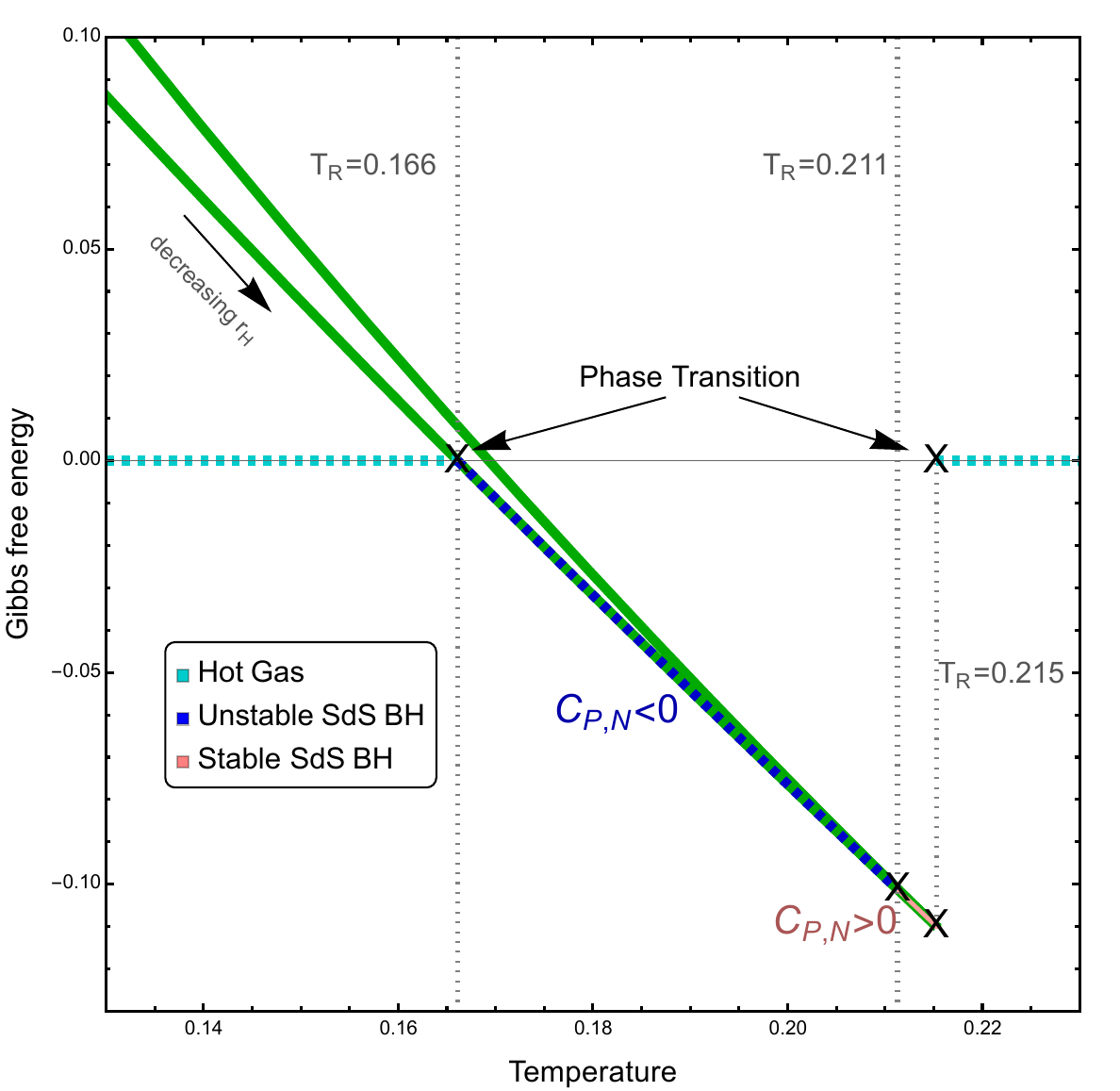}
\caption{Gibbs free energy and Temperature}
\label{fig:stable1}
\end{figure}

In Figure~\ref{fig:stable1}, we show the Gibbs free energy versus temperature under an isobaric process with fixing the number of particles at given temperature represented by the green line. The value of the pressure $\Lambda=0.2$ and the number of particles $N_\lambda=0.3$ are chosen as the same as ones in Figure~\ref{fig:stable}.
The dashed cyan line represents Gibbs free energy of hot gas state. Therefore, there exist the phase transitions from hot gas state to black hole state at the temperature $T_R=0.166$ and $T_R=0.215$. For the hot gas with low temperature $T_R=0.166$, it is 1st order phase transition, and then the system evolves to high temperature until $T_R=0.215$. At this point, the heat capacity is changed its sign corresponding to 2nd phase transition. Since the heat capacity of the black hole in this range $T_R=0.166$ and $T_R=0.215$, is negative, the black hole is locally unstable. The black hole will get higher temperature and smaller radius for this range. Until $T_R=0.215$, a cusp of the graph, the heat capacity is changed its sign corresponding to 2nd order phase transition. After that the system will evolve to a certain temperature suppose to be in the 
range of $r=1.416$ to $r=1.486$ denoted by the pink line as shown in Figure~\ref{fig:stable}. This temperature corresponds to the temperature of the environment and then the black hole are in thermodynamic equilibrium between the black hole and the environment. Note that the pink line in Figure~\ref{fig:stable1} represents the same range with one in Figure~\ref{fig:stable}, corresponding to the range of temperature $T_R=0.211$ to $T_R=0.215$. The $n$th order phase transition is Gibbs free energy diverges at $n$th order differentiation with respect to temperature. \\

\ \ \ For Gibbs-Boltzmann statistics, the heat capacity and the Gibbs free energy can be obtained by taking limitation of $\lambda \rightarrow 0$ to Eq.~\eqref{Cpn}. One found that the temperature taken in the form of Hawking temperature,
\begin{align}
T_H = \lim_{\lambda \rightarrow 0} \left( \frac{\partial m}{\partial S_R} \right)=\frac{\pi - \Lambda S_{BH}}{4 \pi^{3/2} \sqrt{S_{BH}}}.
\end{align}
Since the existence of the black hole horizon is in the range of $r_h < \sqrt{\Lambda}$, or equivalent to $S_{BH} < \frac{\pi}{\sqrt{\Lambda}}$, the temperature is always positive. By considering an isobaric process, the heat capacity is written as 
\begin{align}
C_{\Lambda} = \left. T_H \frac{\delta S_{BH}}{\delta T_H} \right|_{\Lambda} = \left. \frac{T_H}{ \left( \partial T_H /\partial S_{BH} \right) } \right|_\Lambda = - T_H \left( \frac{8 \pi^{3/2} S_{BH}^{3/2}}{\pi + \Lambda S_{BH}} \right) <0 . 
\end{align}
Since $C_\Lambda$ is always negative, the black hole is locally unstable. For global stability, the Gibbs free energy in Gibbs-Boltzmann statistics is obtained by taking $\lambda \rightarrow 0$ to the Gibbs free energy in Eq.~\eqref{Cpn} as follows
\begin{align}
G_{GB} = \lim_{\lambda \to 0} G_{bh} = m - T_H S_{BH}  =\frac{\sqrt{S_{BH}}}{12 \pi^{3/2}} \left( 3\pi + \Lambda S_{BH} \right) >0.
\end{align}
Since $G_{GB}$ is always positive, the black hole is globally unstable. Therefore, the Schwarzschild-de Sitter black hole is thermodynamically unstable based on Gibbs-Boltzmann statistics.

\section{Conclusion \& Discussion} \label{sum}
\ \ \  To study the non-extensive behavior of entropy of a black hole, we have treated Bekenstein-Hawking entropy as the non-extensive Tsallis entropy. In order to describe a black hole by equilibrium thermodynamics, the entropy of a system should be consistent with the 0th law of thermodynamics. The entropy should be additive. R\'enyi entropy is chosen to represent the black hole's entropy as shown in Eq.~\eqref{sbh}. 

In this work, we consider the Schwarzschild-de Sitter black hole which is the solution of Einstein's field equation with the positive cosmological constant, shown in Eq.~\eqref{metric} and \eqref{fr}. Solving the event horizon, the black hole's mass can be written by Eq.~\eqref{mass}. To obtain the relation of black hole's parameters via Smarr's formula, it is necessary to extend the phase space by treating $\Lambda$, and $\lambda$ as variables. As a result, the black hole mass can be written as the homogeneous function of degree $1/2$ in terms of $S_{bh}$, $\Lambda^{-1}$, and $\lambda^{-1}$ as shown in Eq.~\eqref{smarr}. It is reasonable to interpret $\Lambda$ and its conjugate variable as thermodynamic pressure and volume, respectively. For $\lambda$, it and its conjugate variable can be interpreted as the chemical potential and the number of particles, respectively. Since the leading order of its conjugate variable is proportional to the size of the system, corresponding to extensive variable, it is reasonable to interpret its conjugate variable as the number of particles instead of chemical potential.

In Section \ref{cal}, thermodynamic stability of the black hole is investigated by requiring the positive heat capacity and the negative Gibbs free energy. By considering the heat transfer of the black hole as a closed system under an isobaric process. We found that it is possible to obtain a thermodynamically stable black hole for certain range of horizon radius as shown in Figure~\ref{fig:stable}. It is worthwhile to note that the stable black hole cannot be obtained by using Gibbs-Boltzmann description. Moreover, we found that there exists phase transition similar to Hawking-Page phase transition, specifically, a transition between hot gas phase and black hole phase as shown in Figure~\ref{fig:stable1}. For the lower (higher) temperature of phase transition, it undergoes by 1st (0th) order phase transition. 

A Schwarzschild-de Sitter black hole in nature is possibly unstable. If we can detect the stable black hole, the entropy of the black hole should be described by R\'enyi entropy rather than Gibbs-Boltzmann one. The closed thermodynamic system associated with Schwarzschild-de Sitter black hole will be found to be stable under an isobaric process.

There are other types of heat transfer processes, for instance, fixing volume in an open system and a closed system. Therefore, the heat capacities for such the processes, $C_{P,\mu}$, $C_{V,N}$, and $C_{V,\mu}$, should be investigated in order to analyze the stability. On the other hand, heat capacity is a ``response function" that describes how one thermodynamic variable responds to a change of another one under controlled conditions. Since the 1st law of Schwarzschild-de Sitter black hole includes the work term as in found Eq.~\ref{dm}, the Schwarzschild-de Sitter black hole able to respond with a pressure reservoir. Thus, we should consider another response function that corresponds to mechanical expansion which known as ``compressibility". We leave this investigation for further work.

\section{Acknowledgement}
This research has received funding support from the NSF via the Program Management Unit for Human Resources \& Institutional Development, Research and Innovation [ grant number B37G660013 ].

\end{document}